\documentclass[conference, letterpaper]{IEEEtran}
\IEEEoverridecommandlockouts

\usepackage[utf8]{inputenc} 
\usepackage[T1]{fontenc}
\usepackage{url}
\usepackage{ifthen}
\usepackage[cmex10]{amsmath}
\newcommand{\6}{\mathbf }
\usepackage{mathrsfs}
\usepackage{amsfonts}
\usepackage{amsmath}
\usepackage{amssymb}
\usepackage{amsthm}
\usepackage{multirow}
\usepackage{multicol}
\usepackage[dvipsnames]{xcolor}
\usepackage{graphicx}
\usepackage{subcaption}
\usepackage{tikz}
\usepackage{pgfplots}
\usepackage{comment}

\usepackage[ruled, vlined, linesnumbered]{algorithm2e}

\SetCommentSty{mycommfont}

\usepackage[sorting = none, style = numeric-comp, backend = biber, style=ieee]{biblatex}
\newtheoremstyle{bolddefinition}    
  {\topsep}                         
  {\topsep}                         
  {\itshape}                        
  {}                                
  {\bfseries}                       
  {.}                               
  { }                               
  {\thmname{#1}\thmnumber{ #2}\thmnote{ \textbf{(#3)}}} 

\newtheorem{assumption}{Assumption}

\theoremstyle{bolddefinition}
\newtheorem{proposition}{Proposition}[section]

\newtheorem{theorem}{Theorem}[section]

\theoremstyle{bolddefinition}

\newtheorem{lemma}{Lemma}[section]

\newtheorem{remark}{Remark}[section]

\newcommand{\supp}{\mathrm{Supp}}

\pdfmapfile{+custom.map}

\usepackage{mathabx}
\usepackage{pifont}

\usepackage{booktabs}

\newcommand{\bfmax}{\textsf{BF-Max} }

\newcommand{\prob}[1]{\mathrm{Pr}\left[#1\right]}

\usepackage{array}

\pgfplotsset{
    cycle list={
        {red, mark=*},
        {red, mark=x},
        {red, mark=square*},
        {red, mark=triangle*},  {blue,mark=triangle*},
        {red, mark=diamond*},   {blue,mark=diamond*},
        {red, mark=pentagon*},  {blue,mark=pentagon*}
    },
    legend style={
        at={(0.5,-0.2)},
        anchor=north,
        legend columns=2,
        cells={anchor=west},
        font=\footnotesize,
        rounded corners=2pt,
    },
    xlabel=$r$,
    ylabel=$\text{DFR}$,
}

\begin{document}
\title{$\textsf{BF-Max}$: an Efficient Bit Flipping Decoder with Predictable Decoding Failure Rate
\thanks{
This work was partially supported by Agenzia per la Cybersicurezza Nazionale (ACN) under the programme for promotion of XL cycle PhD research in cybersecurity (CUP I32B24001750005), by the Italian Ministry of University and Research (MUR) under the PRIN 2022 program with projects ``Mathematical Primitives for Post Quantum Digital Signatures'' (CUP I53D23006580001) and ``Post quantum Identification and eNcryption primiTives: dEsign and Realization (POINTER)'' (CUP I53D23003670006), by project SERICS (PE00000014) under the MUR National Recovery and Resilience Plan, funded by the European Union - Next Generation EU and by MUR under the Italian Fund for Applied Science (FISA 2022), Call for tender No. 1405 published on 13-09-2022 - project title “Quantum-safe cryptographic tools for the protection of national data and information technology assets” (QSAFEIT) - No. FISA 2022-00618 (CUP I33C24000520001), Grant Assignment Decree no. 15461 adopted on 02.08.2024.
The views expressed are those of the authors and do not represent the funding institutions 
}} 

\author{
  \IEEEauthorblockN{Alessio Baldelli, Marco Baldi, Franco Chiaraluce and Paolo Santini}
  \IEEEauthorblockA{ Dipartimento di Ingegneria dell’Informazione, 
                    Università Politecnica delle Marche, 
                    Ancona, Italy \\
                     \texttt{a.baldelli@pm.univpm.it}, \{\texttt{m.baldi}, \texttt{f.chiaraluce}, \texttt{p.santini}\}\texttt{@univpm.it}}
}

\maketitle


\begin{abstract}
The Bit-Flipping (BF) decoder, thanks to its very low computational complexity, is widely employed in post-quantum cryptographic schemes based on Moderate Density Parity Check codes in which, ultimately, decryption boils down to syndrome decoding. In such a setting, for security concerns, one must guarantee that the Decoding Failure Rate (DFR) is negligible. 
Such a condition, however, is very difficult to guarantee, because simulations are of little help and the decoder performance is difficult to model theoretically.
In this paper, we introduce a new version of the BF decoder, that we call \textsf{BF-Max}, characterized by the fact that in each iteration only one bit (the least reliable) is flipped. 
When the number of iterations is equal to the number of errors to be corrected, we are able to develop a theoretical characterization of the DFR that tightly matches with numerical simulations.
We also show how $\textsf{BF-Max}$ can be implemented efficiently, achieving low complexity and making it inherently constant time.
With our modeling, we are able to accurately predict values of DFR that are remarkably lower than those estimated by applying other approaches.

\end{abstract}

\section{Introduction}
The Bit-Flipping (BF) decoder is probably the simplest decoder for Low-Density Parity-Check (LDPC) codes \cite{gallager1962low}. 
Iterative hard-decision decoders of this family have experienced renewed interest in recent years because they are employed in post-quantum, code-based cryptosystems such as LEDAcrypt \cite{baldi2019ledacrypt} and BIKE \cite{aragon2017bike}. 
In such schemes, decryption requires decoding a syndrome through a Moderate-Density Parity-Check (MPDC) code\footnote{An MDPC code can be thought of as an LDPC code with a somewhat denser parity-check matrix. This is necessary to prevent attacks (such as key recoveries and distinguishers) that exploit the sparsity in the secret key.}.
In these and other applications, BF decoding often is the preferred choice, thanks to its good trade-off between good error correction, low computational complexity and implementation simplicity.
However, as typical in iterative decoding algorithms of this kind, BF is characterized by some nonzero decoding failure rate (DFR), that is, some nonzero probability that decoding fails even for genuine ciphertexts, and decryption failures may leak information about the secret key \cite{guo2016key}.
To avoid such a leakage (formally, to achieve Indistinguishability under Adaptively Chosen Ciphertext Attacks (IND-CCA2)), the DFR must be less than $2^{-\lambda}$, with $\lambda$ being the security parameter (e.g., $\lambda = 128$) \cite{hofheinz2017modular}.
Clearly, such values of DFR cannot be estimated by simulations.

For this reason, in the last few years research has focused on the development of theoretical models to assess the DFR of BF decoders \cite{tillich2018decoding, santini_hard, santini2020analysis, annechini2024bitflipping, sendrier2019decoding, sendrier2020low, baldi2020analysis}.
However, all these works are limited by the need for considering decoders that are somewhat non-optimal, e.g., employ
a very simple flipping criterion (e.g., majority logic decoding \cite{tillich2018decoding}) or a very small number of iterations, say, one as in \cite{tillich2018decoding, santini2020analysis} or two \cite{annechini2024bitflipping}.
Indeed, the lowest DFR is achieved when the decoder runs through many iterations, so that erroneous decisions made by the decoder in the earlier iterations can be revised and corrected in subsequent iterations.
However, this is also what makes estimating the DFR a rather difficult task, because one cannot assume that the errors to be corrected are uniformly distributed at each iteration, and should instead devise a theoretical model that keeps track of their correlation throughout decoding iterations.
To the best of our knowledge, the state of the art in this line of research is the analysis in \cite{annechini2024bitflipping}, where the authors derive upper bounds for a two-iteration, \textit{out-of-place}\footnote{By out-of-place decoder we refer to a decoder that, in each iteration, first flips all the bits it has to flip, and updates the syndrome only after that, before starting the next iteration.
This is opposed to an in-place decoder, which instead updates the syndrome after each bit flip.} BF decoder.

\paragraph*{Our contribution}

In this paper we introduce and analyze $\textsf{BF-Max}$, a BF decoder that flips only one bit per iteration, namely, the one having the lowest reliability. 
The intuition is that, by doing this, the decoder reduces as much as possible the odds that wrong flips happen.
Under standard heuristic assumptions, we are able to derive a closed form formula for the DFR when the number of iterations equals the number of errors. 
Numerical simulations show that the DFR formula we provide is tight and reliable. 
Moreover, we describe a strategy to achieve an efficient  implementation of the \bfmax decoder.
On the negative side, our theoretical model for \bfmax does not take into account the case in which the number of iterations is greater than the number of errors.
Hence, we trade some error correction capability for a closed-form, simple and reliable formula for estimating the DFR.
Still, even with this suboptimal setting, \bfmax outperforms significantly the two-iteration BF decoder studied in \cite{annechini2024bitflipping}: for all parameters considered in \cite{annechini2024bitflipping}, our decoder has comparable time complexity and achieves a significantly lower DFR.
We remark that, by flipping only one bit at each iteration, \bfmax is inherently constant-time.
This, together with the possibility of DFR prediction and its good error correction capability, makes \bfmax an ideal candidate for post-quantum cryptographic schemes based on LDPC and MDPC codes.



\section{Notation and background}
\label{sec:notation}
We denote by $\mathbb F_2$ the binary field.
For clarity, we use different operators for the sum: $+$ and $\sum$ when summing over the reals, and $\oplus$ when summing over $\mathbb F_2$.
Vectors (resp., matrices) over $\mathbb F_2$ are denoted with bold lowercase (resp., uppercase) letters.
The null vector of length $r$ is indicated as $\mathbf{0}_r$.
For a vector $\6a = (a_1,\cdots,a_n)$, we indicate its support as $\supp(\6a)$.
The Hamming weight of $\6{a}$ corresponds to the number of its non-zero entries, and is denoted as $\mathrm{wt}(\6{a})$.
The set of vectors with length $n$ and weight $u$ 
is indicated as $\mathcal S_{n,u}$.
For a set $A$, $x\xleftarrow{\$}A$ indicates that $x$ is sampled uniformly at random from $A$.


\subsection{LDPC codes}

A \textit{linear code} with length $n$ and redundancy $r<n$ is a linear subspace of $\mathbb F_2^n$ with dimension $n-r$.
A code can be represented with a full-rank \textit{parity-check matrix} $\6H\in\mathbb F_2^{r\times n}$, so that the code is the space of all vectors $\6c\in\mathbb F_2^n$ for which  $\6c\6H^\top = \mathbf{0}_r$, where $^\top$ denotes transposition. 
Given a vector $\6e\in\mathbb F_2^n$, its \textit{syndrome} is $\6s = \6e\6H^\top$. If $\6s\neq \mathbf{0}_r$, then $\6e$ is not a codeword; (syndrome) decoding consists in reconstructing $\6e$ from the pair $\{\6H, \6s\}$.
In this paper, we are concerned with LDPC codes,
which are characterized by parity-check matrices having low density, i.e., the number of zero entries in $\6H$ is much less than $rn$.
Among these codes, we focus on those characterized by parity-check matrices having a constant amount $v$ of set entries in each column.

\subsection{Bit-Flipping decoding}

The BF decoder works by iteratively building an estimate for the error vector; such an estimate is initially null and gets refined through an iterative process.
In this process, a key role is played by \textit{counters}: the $i$-th counter, noted by $\sigma_i$, corresponds to the number of unsatisfied parity-check equations in which the $i$-th coordinate participates, namely,
$$\sigma_i = |\left\{j\in\{1,\cdots,r\}\hspace{0.5mm}:\hspace{0.5mm}(h_{j,i} = 1)\wedge(s_j = 1)\right\}|,$$
where $h_{j,i}$ denotes the entry of $\6H$ in the $j$-th row and $i$-th column.
Starting from the computation of counters, it is easy to define the \textit{out-of-place} BF decoding algorithm, which is described in the form of pseudocode in Algorithm \ref{alg:bf}.

We summarize the BF operating principle assuming it receives, as input, a syndrome $\6s = \6e\6H^\top$ with $\6e\in \mathcal S_{n,t}$ and $t$ being properly low.
The decoder either outputs an estimate $\widehat{\6e}\in\mathbb F_2^n$ for the error vector, or $\bot$ whenever decoding fails.
In each iteration, the decoder receives as input the current estimate for the error (which is initially set as the null vector $\mathbf 0_n$) and the syndrome; inside the iteration, the estimate is refined and the syndrome is updated accordingly.
Decisions are taken on the basis of the counters: whenever a counter $\sigma_i$ is high enough (as large as the threshold value for the iteration), it is very likely that the corresponding entry $\widehat e_i$ of the error estimate $\widehat{\6e}$ is wrong and thus gets flipped.
The syndrome is updated accordingly by summing the $i$-th column of $\6H$.
After all updates have been applied, the resulting syndrome is $(\6e\oplus \widehat{\6e})\6H^\top$: if $\6e\oplus \widehat{\6e}$ is a codeword\footnote{We remind that any linear code includes the all-zero vector among its codewords.}, then the syndrome is null and decoding stops, otherwise, another BF iteration starts.
\begin{algorithm}[t]
\small
\KwData{parity-check matrix $\6H\in\mathbb F_2^{r\times n}$, maximum number of iterations $\mathtt{IterMax}\in\mathbb N$, thresholds $(b_1,\cdots, b_{\mathtt{IterMax}})$}
\KwIn{syndrome $\6s \in \mathbb F_2^r$}
\KwOut{decoding failure $\bot$, or vector $\widehat{\6e}\in\mathbb F_2^n$ such that $\mathrm{wt}(\widehat{\6e})\leq \mathtt{IterMax}$ and $\6s = \widehat{\6e}\6H^\top$}
\vspace{2mm}


\tcc{Initialize error estimate and number of performed iterations}
Set $\widehat{\6e} = \mathbf{0}_n$, \quad $\mathtt{Iter} = 1$\; 
\While{$\big(\6s\neq \mathbf{0}_r \big)\vee\big( \mathtt{Iter} \leq \mathtt{IterMax} \big)$}{
Compute counters $(\sigma_1,\cdots,\sigma_n)$\;
\tcc{Update syndrome and error estimate}
\For{$i$ such that $\sigma_i\geq b_{\mathtt{Iter}}$}{
$\6s\gets \6s\oplus \6h_i$;\tcp{$\6h_i$: $i$-th column of $\6H$}
$\widehat e_i\gets \widehat e_i\oplus 1$\;
}
$\mathtt{Iter}\gets \mathtt{Iter}+1$\;
}

\tcc{Output error estimate or report failure}
\textbf{if} $\6s = \mathbf{0}_r$, \Return $\widehat{\6e}$; \textbf{else}, \Return $\bot$\;

\caption{\textsf{BF} decoder \label{alg:bf}
}
\end{algorithm}
If, after the maximum number of iterations has been reached, the syndrome is still not null, then $\widehat{\6e}$ is guaranteed to be wrong and failure is reported.
If, instead, the syndrome is null, then the decoder may have found the true error vector: since the decoder is expected to perform a small number of bit flips, the weight of $\widehat{\6e}$ is expected to be low.
Hence, it is very likely that indeed $\widehat{\6e} = \6e$.

The term \textit{out-of-place} stands for the fact that the syndrome is updated and the counters are recomputed only after all the bit flips of each iteration have been executed; actually, other strategies may be pursued. 
Moreover, there exist variants of BF in which the flipping thresholds are not constant (see e.g. \cite{sendrier2019decoding}).
In this paper, we stick to the choice of constant thresholds, since this is the variant analyzed in \cite{annechini2024bitflipping}.

\paragraph*{Computational complexity} In each iteration, the time complexity is dominated by the cost of computing counters and comparing them with the flipping threshold; exploiting sparsity, which holds, although at different levels, for both LDPC and MDPC codes, this can be done in $O\big(n\cdot (1+v)\cdot \log_2(v)\big)$ operations (here, $\log_2(v)$ accounts for the fact that counters take values in $[0 ; v]$).
Each bit flip requires $O(1+v)$ operations (one for the error update, $v$ for the syndrome update).
Assuming the number of flips is more or less equal to the number of errors to be corrected, we get that, on average, the time complexity of \textsf{BF} is in
$$O\big(\mathtt{IterMax}\cdot n\cdot(v+1)\cdot\log_2(v) + t\cdot (1+v)\big),$$ where, according to Algorithm \ref{alg:bf}, $\mathtt{IterMax}$ represents the maximum number of iterations.
For typical MDPC code parameters, $\mathtt{IterMax}$ is a small constant while both $t$ and $v$ grow as $\sqrt{n}$, resulting in an average complexity $O\big(n^{1.5}\cdot\log_2(n)\big)$.

\section{{$\textsf{BF-Max}$}: efficient implementation}
\label{sec:bf_max_capability}
In this section we introduce the \textsf{BF-Max} decoder and describe how it can be implemented efficiently.

\subsection{Main intuition}

The BF variant we call \bfmax works by flipping a unique bit in each decoding iteration; this bit is selected as the one having the highest counter value among all bits (the case of more counters simultaneously having the highest value is discussed afterwards). 
Remember that, as a rule of thumb, the larger a counter, the higher the probability that the associated bit is wrongly estimated.
By flipping a unique bit, one of those with the largest counter, we “guarantee” that the probability of flipping a wrong bit is reduced to its minimum\footnote{The word guarantee is in quotes since, as we have stressed out, this holds only on the basis of a heuristic reasoning.}.
After the bit is selected, both the error estimate and the syndrome get updated: counters are recomputed and a new iteration starts.

As a little technical caveat, one has to deal with the case in which there are more bits having the largest counter value.         
Many strategies are possible; in this paper, we deal with this case by selecting one of such positions at random.

A straightforward implementation of the \bfmax decoder would recompute all counters in each iteration.  This would lead to a cost of  $\mathtt{IterMax}\cdot \big(n\cdot v\cdot\log_2(v)+v+1\big)$ operations. When there are $t$ errors, we must set $\mathtt{IterMax}\geq t$, which would result in a rather large cost: for typical MDPC code parameters, this would lead to a cost $O\big(n^{2}\cdot\log_2(n)\big)$.
In the next section, we describe how one can instead implement \bfmax with a much lower complexity by exploiting sparsity: for typical MDPC code parameters, we obtain a complexity $O\big(n^{1.5}\cdot\log_2(n)\big)$, which is in the same order of the complexity of the out-of-place BF.

\subsection{Implementation of \bfmax}
We now show how the computational cost of \bfmax can be significantly decreased through an efficient implementation.
The main intuition, here, lies in the observation that,  since \bfmax flips only one bit in each iteration, the number of counters that change, with respect to the ones from the previous iteration, is much less than $n$.
Indeed, every bit participates in exactly $v$ parity-check equations and, in each equation, we have on average $\widebar w = v\cdot n/r$ set entries.
Hence, on average, the number of operations required for updating counters (after one bit flip) is $v \cdot \widebar  w\cdot \log_2(v)$.
Again, considering typical MDPC code parameters, this results in a cost of $O\big(n\cdot \log_2(n)\big)$ operations.
Repeating this for approximately $t = O(\sqrt{n})$ iterations, we get an overall cost of $O\big(n^{1.5}\cdot\log_2(n)\big)$ operations.

Full details about how \bfmax can be implemented with this strategy are given in Algorithm \ref{alg:bfmax_sparse}. 
Counters are computed at the beginning of the process (lines 2--3) and then get updated at the end of each iteration (lines 12--14). 
This is done by considering, for each parity-check equation that changes its value due to the syndrome update (line 11), only the counters corresponding to indices that are in the support of the parity-check equation.
The counter update is either $d = -1$, if the parity-check equation becomes satisfied, or $d = 1$, if the parity-check equation is unsatisfied.
\begin{algorithm*}[!ht]
\small
\KwData{columns supports $\{J_1,\cdots,J_n\}$, rows supports $\{Z_1,\cdots,Z_r\}$, maximum number of iterations $\mathtt{IterMax}\in\mathbb N$}
\KwIn{syndrome $\6s \in \mathbb F_2^r$}
\KwOut{decoding failure $\bot$, or vector $\widehat{\6e}\in\mathbb F_2^n$ such that $\mathrm{wt}(\widehat{\6e})\leq \mathtt{IterMax}$ and $\6s = \widehat{\6e}\6H^\top$}
\vspace{2mm}


Set $\widehat{\6e} = \mathbf 0_n$, \quad $\mathtt{Iter} = 1$;\tcp{Initialize error estimate and number of performed iterations}

\tcc{Initial computation of counters}
\For{$i = 1,\cdots,n$}{ 
Compute $\sigma_i = \sum_{j\in J_i}s_j$\tcp{Integer sum of the syndrome bits indexed by $J_i$}}

\While{$\big(\6s\neq \mathbf{0}_r)\vee\big(\mathtt{Iter}\leq\mathtt{IterMax}\big)$}{

\tcc{Find indices associated to maximum counter value $\widetilde{\sigma}$; store all indices in list $C$}
Set $C = \varnothing$, $\widetilde \sigma = 0$\;
\For{$i = 1,\cdots,n$}{
\textbf{if} $\sigma_i=\widetilde \sigma$, update $C\gets C\cup\{i\}$; \textbf{else}, overwrite $\widetilde \sigma \gets \sigma_i$ and $C \gets \{i\}$\;
}
Set $i^* \xleftarrow{\$} C$;\tcp{Sample at random one of the position with maximum counter}
Update $\widehat e_{i^*}\gets \widehat e_{i^*} \oplus 1$\;

\tcc{Update syndrome and counters: consider only parity-check equations indexed by $J_{i^*}$ (support of column $i^*$), for each equation $j\in J_{i^*}$ update only the counters indexed by $Z_{j}$ (support of row $j$)}
\For{$j\in J_{i^*}$}{
Update $s_j\gets s_j\oplus 1$;\tcp{Update of bit $j$ of the syndrome}
\textbf{if} $s_j = 0$, set $d = -1$; \textbf{else} set $d = 1$;\tcp{$d$ is the counter variation}
\For{$\ell \in Z_j$}{
Update $\sigma_\ell\gets\sigma_\ell + d$;\tcp{Counter update due to row $j$ changing parity}
}
}

$\mathtt{Iter}\gets \mathtt{Iter}+1$;\tcp{Update number of performed iterations}
}


\tcc{Output error estimate or report failure}
\textbf{if} $\6s = \mathbf{0}_r$, \Return $\widehat{\6e}$; \textbf{else}, \Return $\bot$\;

\caption{Efficient implementation of the \textsf{BF-Max} decoder exploiting sparsity\label{alg:bfmax_sparse}
}
\end{algorithm*}

The average time complexity of the algorithm is reported in the next proposition.
\begin{proposition} \label{prop:complexity}
Let $\6H\in\mathbb F_2^{r\times n}$ have constant column weight $v$ and average row weight $\overline w$.
Then, Algorithm \ref{alg:bfmax_sparse} runs in average time
which is well approximated by 
$$n\cdot v\cdot\log_2(v) + \mathtt{IterMax}\cdot\big(n\cdot\log_2(v) + 1 + v + v\overline{w}\big).$$
\end{proposition}
 \begin{IEEEproof}
 The term $n\cdot v\cdot\log_2(v)$ accounts for the initial counters computation.
For each iteration, we neglect some costs (e.g., sampling $i^*$  from $C$) and consider only the following costs: $n \cdot \log_2(v)$ for finding the maximum counter (lines 5--7), $1$ for updating the error estimate and the syndrome (line 9), $v \cdot (1+\overline w)$ for counters and syndrome updates.
Indeed, lines 11--14 are repeated for $v$ times, line 11 takes 1 operation while lines 12--14 take on average $\overline w$ operations (since the average size of $Z_j$ is $\overline w$).
\end{IEEEproof}

\begin{remark}
A non-constant time implementation of a BF decoder would leak side channel information about the secret key \cite{santini2019analysis, eaton2018qc}.
In particular, in the out-of-place BF, the number of bits that are flipped in each iteration is inherently not constant and needs to be properly masked (e.g., by performing a certain number of fake flips): this normally comes with some non-trivial complexity overhead. 
Instead, our decoder is inherently constant time.
Indeed, when working with $(v,w)$-regular LDPC codes, every iteration flips a unique bit and, moreover, takes the same number $v\cdot w\cdot \log_2(v)$ of operations for updating the counters. 
\end{remark}

\section{{$\textsf{BF-Max}$}: modeling the DFR}
\label{sec:bf_max_dfr}
We now describe how the DFR of the \textsf{BF-Max} decoder can be predicted, at least in the case in which the decoder performs a number of iterations exactly equal to the number of errors.

\subsection{DFR prediction}

We first model the probability distribution of the counter values, relying on the following assumption.
\begin{assumption}\label{ass:iid}
Each counter behaves as the sum of independent 
Bernoulli variables, all with the same parameter, which depends only on the value of the corresponding error bit (either 1 or 0).
\end{assumption}

This assumption has been employed in many other papers about BF decoders, at least for what concerns the first decoding iteration (e.g., \cite{santini2019analysis, annechini2024bitflipping, tillich2018decoding}), and is largely accepted when the error positions are uncorrelated.
It is instrumental in deriving the probability distribution for counters  
\cite{santini_hard, annechini2024bitflipping, sendrier2019decoding}.

\begin{proposition}\label{prop:counters_distrib}
Let $\6H\in\mathbb F_2^{r\times n}$ with constant column weight $v$ and row weight $w$.
Let $\6e\xleftarrow{\$} \mathcal S_{n,u}$. 
Under Assumption \ref{ass:iid}, we have
$$\prob{\sigma_i = x\mid  e_i = 1} = g_{1}^{(u)}(x) = \binom{v}{x}\rho_1^x(1-\rho_1)^{v-x},$$
$$\prob{\sigma_i = x\mid e_i = 0} = g_{0}^{(u)}(x) = \binom{v}{x}\rho_0^x(1-\rho_0)^{v-x},$$
where $$\rho_1 = \frac{1}{\binom{n-1}{w-1}}\cdot \sum_{\begin{smallmatrix}\ell = 0\\
\text{$\ell$ \rm{even}}\end{smallmatrix}}^{\min\left\{w-1, u-1\right\}}\binom{u-1}{\ell}\binom{n-u}{w-1-\ell},$$
$$\rho_0 = \frac{1}{\binom{n-1}{w-1}}\cdot \sum_{\begin{smallmatrix}\ell = 1\\
\text{$\ell$ \rm{odd}}\end{smallmatrix}}^{\min\left\{w-1, u\right\}}\binom{u}{\ell}\binom{n-1-u}{w-1-\ell}.$$
\end{proposition} 
We model the DFR of \bfmax using the next assumption.

\begin{assumption}\label{ass:iterations}
Assumption \ref{ass:iid} holds for any iteration of $\bfmax$, if in the previous iterations no wrong bit flip has been performed.
\end{assumption}

Let $i^*$ be the bit which is flipped and initially assume $i^*\not \in E$, where $E$ is the support of the error vector. 
Then, after the first iteration, the vector to be corrected has support $E' = E\cup\{i^*\}$.
While $E$ is chosen uniformly at random among all subsets of $\{1,\cdots,n\}$ of size $t$, this is not true anymore for $E'$: $i^*$ is correlated with the positions in $E$ (since the positions indexed by $E$ caused its flip).
If instead $i^*\in E$, then the error vector which must be corrected in the second iteration has support $E' = E\setminus\{i^*\}$.
$E'$ contains $t-1$ indices which are not correlated, since they have been chosen at the beginning, uniformly at random.
The same reasoning can be repeated for all the subsequent iterations: after iteration $\mathtt{Iter}$, if all the bit flips performed were correct, the number of residual errors is $t-\mathtt{Iter}$ and their positions are uncorrelated.
Thus, for all iterations, Proposition \ref{prop:counters_distrib} is expected to yield valid approximations for the counter distributions. 
In fact, the rationale of considering a number of iterations exactly equal to the number of errors lies in the ability to separate the case in which all flips performed by the decoder are correct from the case in which at least one flipped bit was not affected by error.
In the former case, in fact, the decoder succeeds, while in the latter case it fails, as it cannot perform more iterations than the number of errors to be corrected. 
\begin{proposition}\label{prop:dfr}
Let $\6H\in\mathbb F_2^{r\times n}$ be a parity-check matrix with constant column weight $v$ and constant row weight $w$.
Let $\6e\overset{\$}{\leftarrow}\mathcal S_{n,t}$ and consider \bfmax on input $\6s = \6e\6H^\top$, with $\mathtt{IterMax} = t$.
Then, under Assumptions \ref{ass:iid} and \ref{ass:iterations}, the DFR of \bfmax is well approximated by
$1 - \prod_{u = 1}^t \sum_{x = 0}^{v-1}f_1^{(u)}(x)\cdot f_0^{(u)}(x)$, where $f_1^{(u)}(x) = 
    1 - \left(\sum_{z = 0}^xg_{1}^{(u)}(z)\right)^u$
and
\begin{align*}
f_0^{(u)}(x) = \begin{cases}\hspace{-1mm}\big(g_0^{(u)}(0)\big)^{n-u} & \text{if $x = 0$,} \\
\hspace{-1mm}\left(\displaystyle\sum_{z = 0}^x g_0^{(u)}(z)\right)^{\hspace{-1.5mm}n-u} \hspace{-3.5mm}- \left(\displaystyle\sum_{z = 0}^{x-1} g_0^{(u)}(z)\right)^{\hspace{-1.5mm}n-u} & \text{if $x > 0$.}
\end{cases}
\end{align*}
\end{proposition}
\begin{IEEEproof}
Since $\mathtt{IterMax} = t$, the decoder will not fail if and only if the only bit that flips at each iteration is actually affected by one error.
Thus, after iteration $u\in\{1,\cdots,t\}$, the number of residual errors is $t-u$.
Thanks to Assumption \ref{ass:iterations}, we model the error to be corrected in iteration $u$ as a uniformly random sample from $\mathcal S_{n, t+1-u}$.
For iteration $u$, let $J_0^{(u)}$ and $J_1^{(u)}$ denote the sets of indices of bits in which $\widehat{\6e}$ and $\6e$ are, respectively, equal and different.
Then, in iteration $u$, the decoder will surely take a good choice if
$$\underbrace{\max\left\{\sigma_j\,\left| \,j\in J_0^{(u)}\right.\right\}}_{\widetilde \sigma^{(u)}_0}< \underbrace{\max\left\{\sigma_j\,\left|\, j\in J_1^{(u)}\right.\right\}}_{\widetilde \sigma^{(u)}_1}.$$
Notice that the decoder can still take a good decision, with some probability, even when the above inequality is actually an equality.
For the sake of simplicity, we neglect this possibility (in any case, we expect it happens with low probability).

Then, the DFR can be approximated as
\begin{align*}
& 1 - \prod_{u = 1}^t\prob{\widetilde \sigma_0^{(u)}<\widetilde \sigma_1^{(u)}\,\left| \,\6e\xleftarrow{\$}\mathcal S_{n, t+1-u}\right.}\\\nonumber
& = 1 - \prod_{u = 1}^t\prob{\widetilde \sigma_0^{(u)}<\widetilde \sigma_1^{(u)}\,\left|\, \6e\xleftarrow{\$}\mathcal S_{n,  u}\right.}\\\nonumber
& = 1 - \prod_{u = 1}^t \sum_{x = 0}^{v-1}\prob{ (\widetilde\sigma_0^{(u)} = x)\wedge (\widetilde \sigma_1^{(u)}>x)\,\left|\, \6e\xleftarrow{\$}\mathcal S_{n, u}\right.}.
\end{align*}
Under Assumption \ref{ass:iid}, the counters are independent random variables, hence
\begin{align*}
&\prob{(\widetilde\sigma_0^{(u)} = x)\wedge (\widetilde \sigma_1^{(u)}>x)\,\left|\, \6e\xleftarrow{\$}\mathcal S_{n, u}\right.}\\\nonumber
&\quad = \underbrace{\prob{\widetilde\sigma_0^{(u)} = x\,\left|\, \6e\xleftarrow{\$}\mathcal S_{n, u}\right.}}_{f_0^{(u)}(x)}\cdot\underbrace{\prob{\widetilde\sigma_1^{(u)} > x\left|\, \6e\xleftarrow{\$}\mathcal S_{n, u}\right.}}_{f_1^{(u)}(x)}.
\end{align*}
With further probability theory arguments and recalling Proposition \ref{prop:counters_distrib}, we obtain the expressions for $f_0^{(u)}(x)$ and $f_1^{(u)}(x)$.
\end{IEEEproof}


\subsection{Numerical results}

\begin{figure}[t!]
\centering
\resizebox{\columnwidth}{!}{
\input{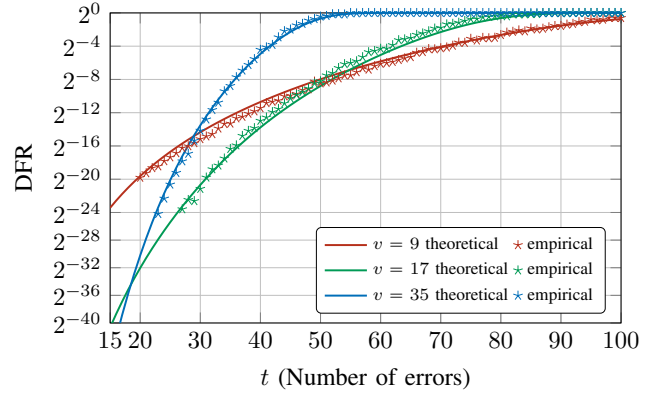}
}
\caption{Comparison between theoretical and empirical DFR of the \bfmax decoder, for QC-LDPC 
codes with fixed $r = 2003$, $n = 2r$, $w = 2v$ and several values of $v$.
}
\label{fig:Our_QC_DFR}
\end{figure}

\begin{figure}
\centering
\resizebox{\columnwidth}{!}{
\input{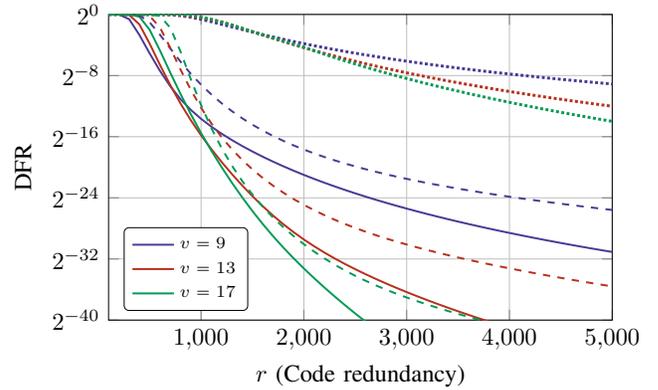}
}
\caption{Comparison between the DFR theoretical models for \bfmax (continuous lines), 2-iteration BF (dashed lines) and 1-iteration BF (dotted lines).
For all curves, $t = 18$, $w = 2v$ and $n = 2r$; the flipping thresholds for BF have been optimized in order to achieve the lowest DFR. The same color has been used for curves referred to the same tuple $(t, v, w)$.
}
\label{fig:comparison}
\end{figure}

To validate Proposition \ref{prop:dfr}, we can compare it with estimates obtained by Monte Carlo simulations.
In Fig.~\ref{fig:Our_QC_DFR}, we compare the DFR of some quasi-cyclic (QC) codes predicted by Proposition \ref{prop:dfr}, with the one estimated through numerical simulations \footnote{The code employed for the simulations, as well as an implementation for the formula in Proposition \ref{prop:dfr}, can be found at \url{https://github.com/secomms/bf-max}.}.
As in BIKE and LEDAcrypt, we consider codes whose parity-check matrix is in the form $(\6H_1, \6H_2)$, with both $\6H_1$ and $\6H_2$ being circulant matrices with size $r \times r$ and column weight $v$.
The resulting parity-check matrix has $n = 2r$ and constant row weight $w = 2v$.
As can be seen from Fig.~\ref{fig:Our_QC_DFR}, the results of the simulations are aligned with the theoretical predictions.
In Fig. \ref{fig:comparison} we compare the DFR resulting from Proposition \ref{prop:dfr} with the theoretical models for the \textsf{BF} decoder, considering one and two iterations \cite{santini_hard, annechini2024bitflipping}.
As we can see, our decoder has a much lower DFR.
Moreover, all decoders have comparable computational complexity: for instance, for $v = 9$, the complexity of \bfmax is approximately 30\% greater than that of the 2-iterations \textsf{BF}, while for $v = 17$ they have essentially the same complexity.

\section{Conclusion}
\label{sec:conclusions}
We have introduced and analyzed the \bfmax decoder, a BF decoder that flips a unique bit in each iteration choosing that (or one of those) with the highest counter.
We have shown that this decoder has low computational complexity and is inherently constant time.
We have been able to characterize its DFR theoretically, for the case in which the number of iterations equals the number of errors to be corrected.
The DFR predicted through our model closely matches the results of numerical simulations and significantly outperforms the theoretical DFR obtained by other models for BF decoders.


\printbibliography

@inproceedings{santini2019analysis,
  title={Analysis of reaction and timing attacks against cryptosystems based on sparse parity-check codes},
  author={Santini, Paolo and Battaglioni, Massimo and Chiaraluce, Franco and Baldi, Marco},
  booktitle={Proc. Code-Based Cryptography - 7th International Workshop (CBC 2019), Revised Selected Papers},
  address={Darmstadt, Germany},
  pages={115--136},
  month={May},
  year={2019},
  organization={Springer}
}

@inproceedings{eaton2018qc,
  title={{QC-MDPC}: a timing attack and a {CCA2 KEM}},
  author={Eaton, Edward and Lequesne, Matthieu and Parent, Alex and Sendrier, Nicolas},
  booktitle={Proc. International Conference on Post-Quantum Cryptography},
  address={Fort Lauderdale, Florida},
  pages={47--76},
  month={Apr.},
  year={2018}
}

@inproceedings{tillich2018decoding,
  title={The decoding failure probability of {MDPC} codes},
  author={Tillich, Jean-Pierre},
  booktitle={Proc. 2018 IEEE International Symposium on Information Theory (ISIT)},
  pages={941--945},
  address={Vail, CO},
  month={Jun.},
  year={2018}
}

@inproceedings{baldi2020analysis,
  title={Analysis of in-place randomized bit-flipping decoders for the design of {LDPC} and {MDPC} code-based cryptosystems},
  author={Baldi, Marco and Barenghi, Alessandro and Chiaraluce, Franco and Pelosi, Gerardo and Santini, Paolo},
  booktitle={Proc. 17th International Conference on E-Business and Telecommunications (ICETE 2020)},
  address={Virtual, Online},
  pages={151--174},
  month={Jul,},
  year={2020}
}

@inproceedings{hofheinz2017modular,
  title={A modular analysis of the {F}ujisaki-{O}kamoto transformation},
  author={Hofheinz, Dennis and H{\"o}velmanns, Kathrin and Kiltz, Eike},
  booktitle={Proc. 15th Theory of Cryptography Conference (TCC ’17)},
  address={Baltimore, MD},
  pages={341--371},
  month={Nov.},
  year={2017}
}

@inproceedings{guo2016key,
  title={A key recovery attack on {MDPC} with {CCA} security using decoding errors},
  author={Guo, Qian and Johansson, Thomas and Stankovski, Paul},
  booktitle={Proc. Advances in Cryptology--ASIACRYPT 2016: 22nd International Conference on the Theory and Application of Cryptology and Information Security},
  address={Hanoi, Vietnam},
  pages={789--815},
  month = {Dec.},
  year={2016}
}

@article{baldi2019ledacrypt,
  title={{LEDA}crypt},
  author={Baldi, Marco and Barenghi, Alessandro and Chiaraluce, Franco and Pelosi, Gerardo and Santini, Paolo},
  journal={Second round submission to the NIST post-quantum cryptography call},
  year={2019}
}

@article{aragon2017bike,
  title={{BIKE}},
  author={Aragon, N and Barreto, P and Bettaieb, S and Bidoux, Loic and Blazy, O and Deneuville, JC and Gaborit, P and Gueron, S and G{\"u}neysu, T and Melchor, C Aguilar and others},
  journal={First round submission to the NIST post-quantum cryptography call},
  year={2017}
}

@inproceedings{sendrier2019decoding,
  title={On the decoding failure rate of {QC-MDPC} bit-flipping decoders},
  author={Sendrier, Nicolas and Vasseur, Valentin},
  booktitle={Post-Quantum Cryptography - 10th International Conference (PQCrypto 2019) Revised Selected Papers},
  address={Chongqing, China},
  pages={404--416},
  month={May},
  year={2019}
}

@inproceedings{sendrier2020low,
  title={About low {DFR} for {QC-MDPC} decoding},
  author={Sendrier, Nicolas and Vasseur, Valentin},
  booktitle={Proc. International Conference on Post-Quantum Cryptography},
  address={Paris, France},
  pages={20--34},
  month={Sep.},
  year={2020}
}

@article{gallager1962low,
  title={Low-density parity-check codes},
  author={Gallager, Robert},
  journal={IRE Transactions on information theory},
  volume={8},
  number={1},
  pages={21--28},
  year={1962},
  publisher={IEEE}
}

@article{santini2020analysis,
  title={Analysis of the error correction capability of {LDPC} and {MDPC} codes under parallel bit-flipping decoding and application to cryptography},
  author={Santini, Paolo and Battaglioni, Massimo and Baldi, Marco and Chiaraluce, Franco},
  journal={IEEE Transactions on Communications},
  volume={68},
  number={8},
  pages={4648--4660},
  year={2020},
  publisher={IEEE}
}

@INPROCEEDINGS{santini_hard,  author={Santini, Paolo and Battaglioni, Massimo and Baldi, Marco and Chiaraluce, Franco},  booktitle={Proc. 2019 IEEE International Conference on Communications (ICC)},   title={Hard-Decision Iterative Decoding of {LDPC} Codes with Bounded Error Rate},
address={Shanghai, China},  month={May}, year={2019}}

@inproceedings{annechini2024bitflipping,
  title={Bit-flipping Decoder Failure Rate Estimation for (v,w)-regular Codes},
  author={Alessandro Annechini and Alessandro Barenghi and Gerardo Pelosi},
  booktitle={Proc. 2024 IEEE International Symposium on Information Theory (ISIT)},
  pages={3375--3379},
  address={Athens, Greece},
  month={Jul.},
  year={2024}
}

\end{document}